\documentclass[twocolumn]{article}

\usepackage{filecontents}
\makeatletter\@input{z1.tex}\makeatother
\makeatletter\@input{z2.tex}\makeatother
\makeatletter\@input{z3.tex}\makeatother

\usepackage[paperwidth=210mm,paperheight=297mm,centering,hmargin=2cm,vmargin=2.5cm]{geometry}

\usepackage{authblk}

\usepackage{abstract}

\usepackage[numbers,sort&compress]{natbib}
\usepackage{multibib}
\newcites{methods}{Methods References}

\usepackage{titlesec}
\titleformat{\subsection}  
  {\fontsize{10}{15}\sffamily} 
  {\S\thesubsection} 
  {2em} 
  {\bfseries \centering} 
  [] 

\DeclareUnicodeCharacter{25CF}{$\bullet$}
\usepackage{graphicx}
\usepackage[dvipsnames]{xcolor}
\usepackage{pdfpages}
\usepackage{amssymb}
\usepackage{amsmath}
\usepackage{amsbsy}
\usepackage{color}
\usepackage{epstopdf}
\usepackage{bm}
\usepackage{xurl}

\usepackage[hidelinks]{hyperref}
\usepackage{textcomp, gensymb}
\usepackage[normalem]{ulem}

\graphicspath{{img/}}

\usepackage{xr}          
\makeatletter
\AtBeginDocument{\let\LS@rot\@undefined}
\newcommand*{\addFileDependency}[1]{
  \typeout{(#1)}
  \@addtofilelist{#1}
  \IfFileExists{#1}{}{\typeout{No file #1.}}
}
\makeatother
\newcommand*{\myexternaldocument}[1]{%
    \externaldocument[SM-]{#1}%
    \addFileDependency{#1.tex}%
    \addFileDependency{#1.aux}%
}
\myexternaldocument{SI}

\title{\vspace{-2.0cm} Topological Linking Determines Elasticity in Limited Valence Networks}
\author[1]{Giorgia Palombo}
\author[1]{Simon Weir}
\author[1,2,*]{Davide Michieletto}
\author[1,$\dagger$]{Yair Augusto Guti\'{e}rrez Fosado}
\affil[1]{School of Physics and Astronomy, University of Edinburgh, Peter Guthrie Tait Road, Edinburgh, EH9 3FD, UK}
\affil[2]{MRC Human Genetics Unit, Institute of Genetics and Cancer, University of Edinburgh, Edinburgh EH4 2XU, UK}
\affil[*]{\footnotesize{corresponding author: davide.michieletto@ed.ac.uk}}
\affil[$\dagger$]{\footnotesize{corresponding author: yair.fosado@ed.ac.uk}}

\date{} 

\begin{document}

\twocolumn[
  \begin{@twocolumnfalse}
  \maketitle
\begin{abstract}
\vspace{-1.0cm}
\textbf{Understanding the relationship between the microscopic structure and topology of a material and its macroscopic properties is a fundamental challenge across a wide range of systems. Here, we investigate the viscoelasticity of DNA nanostar hydrogels -- a model system for physical networks with limited valence -- by coupling rheology measurements, confocal imaging and molecular dynamics simulations. We discover that these networks display a large degree of interpenetration and that loops within the network are topologically linked, forming a percolating network-within-network structure. Below overlapping concentration, the fraction of branching points and the pore size determine the high-frequency elasticity of these physical gels. At higher concentrations, we discover that this elastic response is dictated by the abundance of topological links between looped motifs in the gel. Our findings highlight the emergence of ``topological elasticity'' as a previously overlooked mechanism in generic network-forming liquids and gels and inform the design of topologically-controllable material behaviours.
}
\end{abstract}
\end{@twocolumnfalse}
]

Networks made by colloids and polymers are fundamental components of virtually any material around us~\cite{degennes_scaling}, and even a substance as familiar as water exhibits complex network structures~\cite{Sciortino2022,Fosado2024ice}. Understanding the relationship between the microscopic structure of these networks and their meso/macroscopic material properties has been the focus of the material science community for decades. Classic textbook pictures mainly focus on entanglements~\cite{Everaers2004} and  crosslinks~\cite{degennes_scaling} to predict the material properties of gels and complex fluids; however, the role played by network motifs that impose formal topological invariants is far less understood~\cite{Tubiana2024}.

Most classic theories rely on chemical composition, fraction of branching points, mesh size and fractal dimension to predict the mechanical properties of soft colloidal and polymeric gels~\cite{Flory1953}. However, it is becoming apparent that to accurately describe a material's behaviour, we also require a quantitative understanding of rigorously-defined topological motifs created by the microscopic building blocks, such as loops, knots and links. This need is especially evident in, e.g., gels with looped defects~\cite{Gu2018,Gu2019,Lin2018}, solutions of ring polymers~\cite{Michieletto2016pnas,Kapnistos2008,Smrek2021,OConnor2020}, Olympic gels~\cite{Krajina2018,Lang2014a,He2022}, molecular knots~\cite{Leigh2020}, polyrotaxanes~\cite{Imran2014}, soft particulate gels~\cite{smith2024}, and even associating liquids, such as water~\cite{Sciortino2022,Fosado2024ice}.

Especially challenging are physical networks with limited valence, where the connectivity of the building blocks is constrained and are thus expected to display unconventional network structures~\cite{Sciortino2022,Frank2013,ConradPNAS2019,Sharma2016}. 
Experimentally, limited valence networks are most commonly found in associating liquids~\cite{Sciortino2022}, but are best investigated ``at larger time- and length-scales'' through model systems made of patchy colloids~\cite{ChenQian2011,WangYu2012}, or DNA nanostars~\cite{SciortinoPNAS2013}, where single-stranded DNA oligomers are self-assembled into star-shaped building blocks with a fixed number of arms and terminal ``sticky ends'' forming reversible (or physical) networks~\cite{SciortinoPNAS2013, BiffiSM2015,EiserPNAS2018}. DNA nanostars (DNAns) self-assemble into complex fluids and physical gels with a range of viscoelastic properties, which have been studied extensively in recent years~\cite{SciortinoPNAS2013, EiserPNAS2018, ConradPNAS2019}. 

\begin{figure*}[t!]
	\centering
	\includegraphics[width=1.0\textwidth]{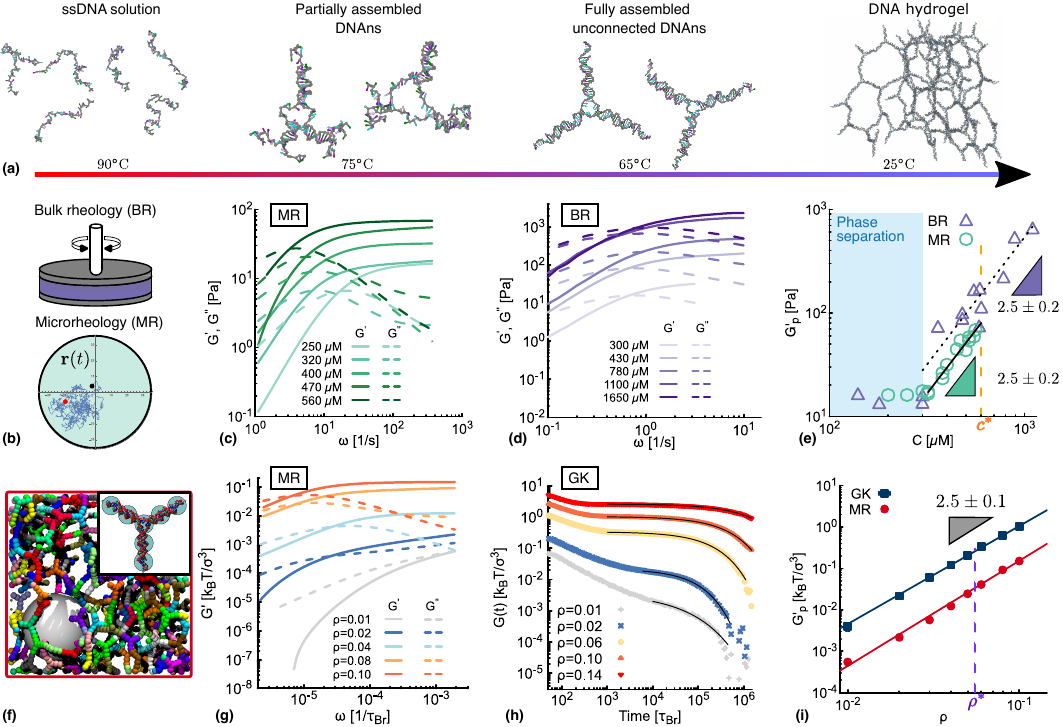}
\caption{\textbf{Scaling of elasticity in gels of DNA nanostars}.  \textbf{(a)} Schematic of the annealing procedure for assembling individual DNAns and the full hydrogel. \textbf{(b)-(e)} Bulk rheology (BR) and microrheology (MR) results at different DNAns concentrations. \textbf{(c)-(d)} Elastic ($G^{\prime}(\omega)$) and viscous ($G^{\prime\prime}(\omega)$) moduli from MR and BR, respectively. \textbf{(e)} Elastic plateau $G^{\prime}_{p}$ as function of DNAns concentration (see also Supplementary Fig.~\ref{SM-fig:BR} with errors in the concentration of DNAns). The cyan-shaded area represents the concentrations at which the system phase separates~\cite{Saleh2022}. Data is well fitted by a power law $G^{\prime}_{p} \sim C^{2.5}$ in the gel region for both techniques (black lines). Labels give best-fit exponents with standard fitting error estimates. Orange dashed line indicates the overlapping concentration $c^{*}=600\mu$M (see Supplementary Section~\ref{SM-sec:overlapConc}). \textbf{(f)} Schematic of the simulated coarse-grained DNAns hydrogel and its assembly into a network. A large bead (white) is used to simulate MR. \textbf{(g)} $G^{\prime}$ and $G^{\prime\prime}$ from MR simulations. \textbf{(h)} Stress-relaxation function from equilibrium Green-Kubo (GK) simulations. Black lines are fits of a stretched exponential function to the data (see Supplementary Fig.~\ref{SM-fig:GK} for details). \textbf{(i)} Elastic plateau ($G^{\prime}_{p}$) as a function of the volume fraction $\rho$ from simulated MR and the GK relation (see Supplementary Section~\ref{SM-sec:elastSim} for details). Purple dashed line indicates the overlapping volume fraction $\rho^{*}=0.056$. Data points represent mean $\pm$ SEM (smaller than symbol size) from five independent replicates. Data are well fitted by a power law $G^{\prime}_{p} \sim \rho^{2.5}$ in the whole range of concentrations.}
    \label{fig:experiments}
\end{figure*}

We consider DNAns gels as model systems to study generic networks formed by limited valence building blocks (see Fig.~\ref{fig:experiments}(a)). Although the operational definition of a gel is that its elastic response dominates at arbitrarily low-frequencies, here we adopt the terminology used in the literature~\cite{ConradPNAS2019,EiserPNAS2018,SciortinoPNAS2013}. We thus refer to ``DNAns gels'' and to their elasticity, specifically in the context of the high frequency regime, where their behaviour is elastic-dominated. DNAns gels are in fact complex fluids that display a Maxwellian behaviour, i.e., elastic at large frequencies but liquid at long enough timescales~\cite{SciortinoPNAS2013,Rovigatti2014,EiserPNAS2018,ConradPNAS2019}. Gaining a better understanding of DNAns hydrogel rheology will speed up the bottom-up discovery and formulation of new DNA-based soft materials.

We couple extensive molecular dynamics (MD) simulations of an oxDNA-inferred~\cite{FosadoSM2023} coarse-grained model of DNAns, with bulk and microrheology experiments and confocal imaging, to elucidate the connection between network structure and, more importantly, topology (such as knots and links), and its elastic behaviour. By spanning a range of concentrations $C$, we discover that the system undergoes a topological transition around the overlapping concentration $c^{*}$, where loops within the network start to link with each other. While for $C<c^{*}$ the fraction of branching points and the mesh size of the network , i.e. its \textit{structure}, govern the elasticity of the gel, for $C \geq c^{*}$ linking of loops within the network, i.e. its \textit{topology}, emerges as the main determinant of the gels' mechanical properties. In the latter regime, the elasticity of the gel, i.e. its solid-like response to high frequency stress, scales with the concentration of DNAns as $G^{\prime}_p \sim C^{2.5}$ and linearly with the linking number between minimum loops $G^{\prime}_p \sim \mathcal{L}$. 

This simple relationship connects, for the first time to our knowledge, the macroscopic mechanical properties of a soft material with a mathematically rigorous topological invariant which is markedly distinct from generic ``entanglements'' or crosslinks. 

\subsection*{\centering Scaling of elasticity in physical gels with limited valence} 

We first assembled tri-armed DNA nanostars as done previously (see Supplementary Section~\ref{SM-sec:nsdesign} and Ref.~\cite{SciortinoPNAS2013,ConradPNAS2019}). We then prepared solutions at different concentrations by annealing the DNAns via quenching from 90$^\circ$C to 25$^\circ$C (see Fig.~\ref{fig:experiments}(a) and Supplementary Section~\ref{SM-sec:hydrogelprep}). Particle tracking microrheology (MR) and oscillatory bulk rheology (BR) were then employed to measure the frequency-dependent elastic ($G^{\prime}(\omega)$) and viscous ($G^{\prime\prime}(\omega)$) modulus, see Fig.~\ref{fig:experiments}(c),(d) and Supplementary Section~\ref{SM-sec:expProtocol}. In the Supplementary Fig.~\ref{SM-fig:BR}, we also show that $G^{\prime}$ and $G^{\prime\prime}$ from different concentrations superimpose into a master curve that follows a near-Maxwellian behaviour with a characteristic timescale $\sim \omega_0^{-1}$, due to the reversible nature of the hybridisation between DNAns.

We extracted the elastic plateau, $G^{\prime}_{p}$, as the value of the elastic modulus at the largest frequency $\omega \simeq 10^{2}$ Hz (although we obtained similar results by choosing the value of $G^{\prime}$ at the crossover point, see Supplementary Section \ref{SM-sec:BR}). The measurements from BR and MR are in very good agreement, yielding a scaling $G^{\prime}_{p}\sim C^{2.5\pm0.2}$ at $C>300\mu$M (beyond the phase-separation region, see Fig~\ref{fig:experiments}(e)), in agreement with values obtained for other limited-valence gels~\cite{Dudukovic2014}. 
The values of $G^{\prime}_p$ in BR are consistently larger than the ones found in MR, likely due to surface effects between the DNA and the probe particle~\cite{SchmidtBRMR} (see Supplementary~Fig.~\ref{SM-fig:BRT25}). Whilst the scaling exponent found here is larger than the one reported in the literature for a similar system, both the range of elasticity ($G^{\prime}_{p} \sim 10 - 700$ Pa) and relaxation timescales ($\tau_{u} \sim 1$ s) are in line with that of Ref.~\cite{ConradPNAS2019}. To better understand how this scaling emerges from the network structure, we turned to MD simulations.

\subsection*{\centering Coarse-grained model captures bulk behaviour of DNA networks}

We used a recently developed computational coarse-grained model of rigid analogues of DNAns~\cite{FosadoSM2023} with valence $f=3$ (Fig.~\ref{fig:experiments}(f)). Briefly, each nanostar is modelled as a rigid body with a Y-shaped structure that was inferred from oxDNA simulations~\cite{oxDNALAMMPS,FosadoSM2023} (see Methods and Supplementary Section~\ref{SM-sec:simDetails}). We simulate solutions of DNAns at different volume fractions, ranging from $\rho=0.01$ to $0.14$. To measure the viscoelasticity, we first track the position of a large bead embedded in the sample and compute its mean squared displacement $\text{MSD}(t)= \langle \lvert \mathbf{r}(t+t_{0})-\mathbf{r}(t_{0}) \rvert^{2} \rangle$, where the average is performed over different values of $t_{0}$. We use the generalized Stokes-Einstein relation to compute the complex stress modulus~\cite{Mason2000Gp} and the elastic plateau $G^{\prime}_{p}$ (see Fig.~\ref{fig:experiments}(g) and Supplementary Section~\ref{SM-sec:elastSim}). In parallel, we also use the Green–Kubo relation and compute the auto-correlation of the off-diagonal components of the stress-tensor to obtain the stress relaxation function~\cite{Ramirez2010,Lee2009} $G(t) = \frac{L^3}{3k_{B}T} \sum_{\alpha \neq \beta} P_{\alpha\beta}(0) P_{\alpha\beta}(t)$  ($P_{\alpha\beta}=P_{xy}$, $P_{xz}$ and $P_{yz}$, see Methods for details). 
The elastic plateau of the system is then obtained by fitting a stretched exponential function $f(t)=a \exp{(-(t/\tau)^{b})}$, in which $a$ and $\tau$ represent the elastic plateau and the relaxation time of the network, respectively~\cite{Flenner2015} (see Fig.~\ref{fig:experiments}(h)). 
Both methods are in excellent agreement and yield a network elasticity that scales as $G^{\prime}_p \sim \rho^{2.5 \pm 0.1}$ (Fig.~\ref{fig:experiments}(i)), exactly matching the one found in our experiments. In fact, we even obtain the same result with a more flexible DNAns model (see Supplementary Fig.~\ref{SM-fig:nonRigid}). Thus, we argue that our simulations well capture the behaviour and internal microstructure of the gels. In line with Ref.~\cite{ConradPNAS2019}, our simulations suggest that the gels' elasticity displays a non-linear scaling with concentration different from predictions of classic models, such as phantom network theory. To understand the underlying physical origin of elasticity in these networks, we thus now investigate the microscopic properties and topology of our simulated DNAns gels.

\begin{figure*}[t!]
	\centering
	\includegraphics[width=0.9\textwidth]{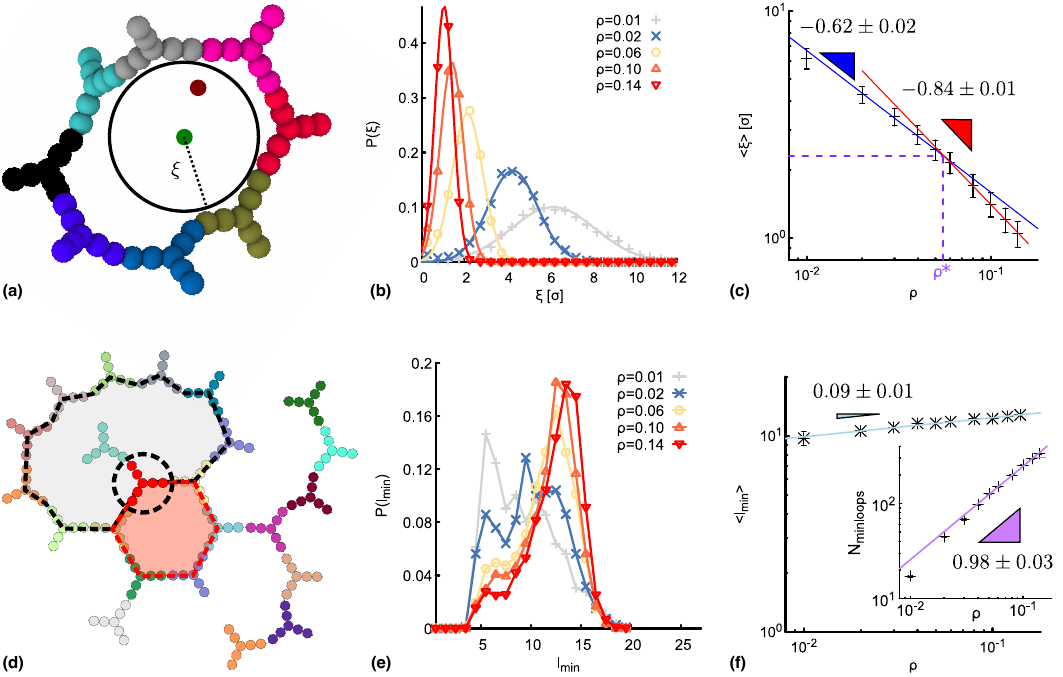}
	\caption{\textbf{Characterization of geometrical features in simulated DNAns networks.} \textbf{(a)} Schematic of the random insertion method~\cite{Sorichetti2020}: the largest sphere with radius $\xi$ containing a random point (in red) is shown. \textbf{(b)} Probability distribution of the mesh size $\xi$. Lines are fits to the data using a Gaussian function. \textbf{(c)} Mean and SEM of the mesh size obtained from the previous fit, using 50 independent gel configurations per volume fraction. \textbf{(d)} Sketch of the minimum loop search algorithm. We highlight the two loops passing through the red DNA nanostar (shaded grey and red). The minimum loop, with $l_{min}=6$ nanostars is shaded red. \textbf{(e)} Probability distribution of the number of nanostars in the minimum loops $l_{min}$. Lines are for visual guidance. The mean and SEM of the distribution at different concentrations are used to show the $l_{min}$ scaling with $\rho$ in panel \textbf{(f)}. Inset shows scaling of the number of minimum loops with volume fraction. Error bars are smaller than symbols. Results in this panel use at least 25 independent gel configurations per volume fraction. See Supplementary Section~\ref{SM-sec:TopaAnalysis}, Figs.~\ref{SM-fig:spath} and \ref{SM-fig:minloop} for more details and scaling of distance between branch points.}
	\label{fig:NetworkGeometry}
\end{figure*}

\subsection*{\centering Length scales in DNAns gels suggest network interpenetration}

One of the most important length scales in polymers and colloidal solutions is the mesh size ($\xi$), or the size of the gel pores. We computed the mesh size in our simulated gels by a method of random insertions~\cite{Sorichetti2020}, done by randomly choosing a point in our simulation box and by finding the largest sphere that satisfies two conditions: containing the point and not touching any of the nanostar beads (Fig.~\ref{fig:NetworkGeometry}(a)). We repeat this procedure $200,000$ times over $50$ independent gel configurations and obtain the probability distribution ($P(\xi)$) of the mesh size, i.e. the diameter of the space-filling sphere, at different volume fractions (see Fig.~\ref{fig:NetworkGeometry}(b)). We find Gaussian-distributed values, from which we extract the average mesh size $\langle \xi \rangle$ (Fig.~\ref{fig:NetworkGeometry}(c)). Interestingly, at the overlap concentration, $\rho^{*}=0.056$, we identify a transition where the scaling of $\langle \xi \rangle$ changes from $\xi \sim \rho^{-0.62}$ to $\xi \sim \rho^{-0.84}$ (see Supplementary Section \ref{SM-sec:overlapConc}). We interpret this as a sign of a structural transition suggesting the onset of a network geometry that allows for tighter packing. This length scale appears to have a similar scaling to that measured in strain hardening experiments~\cite{ConradPNAS2019}.

Another key structural feature in gels is the presence of loops within the network. While loops are often associated with weak motifs (or defects) in polymer networks as they reduce the density of ``active'' crosslinks~\cite{Zhong2016}, we argue that in the case of branched, amorphous networks, short loops may also act as topological springs. To quantify the abundance of loops within the network, we employed graph analysis (see Supplementary Section~\ref{SM-sec:TopaAnalysis}), focusing on so-called ``minimum'' loops, which are computed as follows: for each DNAns (e.g., the one circled with dashed lines in Fig.~\ref{fig:NetworkGeometry}(d)) we found all the loops passing through it and selected the loop with the least number of DNAns (highlighted by the red shaded area in Fig.~\ref{fig:NetworkGeometry}(d)). We then compiled the statistics of all minimum loops in the network and calculated the probability $P(l_{min})$ that a minimum loop is formed by $l_{min}$ nanostars and its first momentum $\langle l_{min} \rangle$ (Fig.~\ref{fig:NetworkGeometry}(e,f)), which we found has values similar to the estimated cluster size in Ref.~\cite{ConradPNAS2019}. Remarkably, these distributions shift to larger values of $\langle l_{min} \rangle$ for larger concentrations of DNAns, indicating that the average minimum loop length \textit{increases} with DNAns concentration. This is in marked contrast with the finding above, i.e. that the mesh size decreases with DNAns concentration. Indeed, albeit the scaling of $l_{min}$ is weak, $\langle l_{min} \rangle \sim \rho^{0.1}$, the number of minimum loops is extensive in $\rho$ (Fig.~\ref{fig:NetworkGeometry}(f), inset).

Additional quantities that typically characterise the mechanical properties of gels are (i) the number of branching points and (ii) the distribution of the shortest path ($\lambda$) connecting two branching points, $P(\lambda)$~\cite{Delgado2023}. In marked contrast with typical results from chemical cross-linking gels, we discover that the average path length $\langle \lambda \rangle$ \textit{increases} with DNAns concentration, with a similar exponent found for the growth of minimum loop length (see Supplementary Section~\ref{SM-sec:TopaAnalysis}). Importantly, we also find that the fractal dimension of the gel ($d_{f}=1/\nu=1.7$) is independent of the concentration of DNAns, suggesting that the transition from liquid to gel behaviour must be accompanied by a distinct type of transition in the network that does not affect the geometric and fractal arrangement of the nanostars (see Supplementary Fig.~\ref{SM-fig:spath}).

\begin{figure*}[t!]
	\centering
	\includegraphics[width=0.85\textwidth]{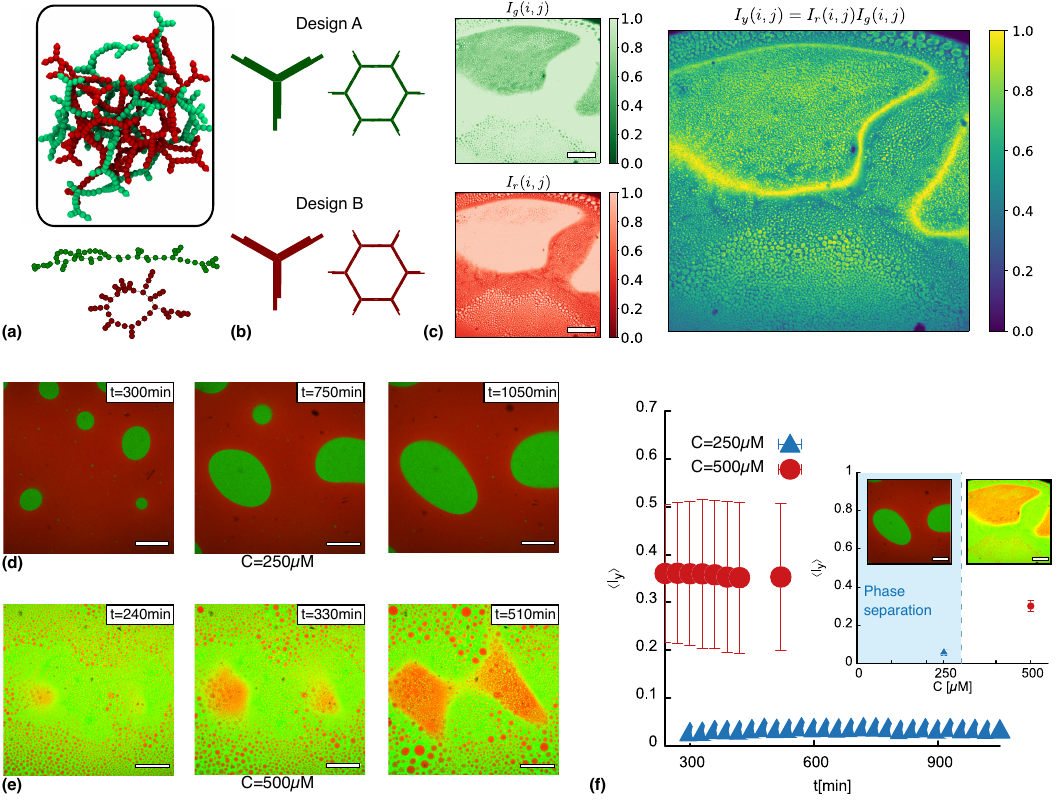}
	\caption{\textbf{Interpenetration of DNAns networks.} \textbf{(a)} Snapshot from simulations performed with only one type of DNAns at $\rho=0.04$ displaying interpenetrated structures in a 3D spherical neighbourhood of the simulation box, with the corresponding disjoint graph diagrams at the bottom (we found similar results in simulations of a binary mixture of DNAns, see Supplementary Fig.~\ref{SM-fig:mixedAB}). \textbf{(b)} Schematic representation of DNA nanostars designs for the experimental interpenetration assay. The cores of DNS-A (green) and DNS-B (red) have the same nucleotide composition but variations in their sequence that avoid accidental cross-hybridisation. The two designs have orthogonal overhangs with the same lengths and interaction strengths, allowing for A-A and B-B interactions but no A-B hybridisation. \textbf{(c)} Confocal images showing partial interpenetration when mixing [DNS-A]=500$\mu$M and [DNS-B]=500$\mu$M. In the left panel, we show images in the two channels: FAMK (488 nm green, top) and Cy3 (555 nm, red, bottom). The right panel is the pixel-by-pixel product of the images, or yellow signal, $I_y(i,j) = I_r(i,j) I_g(i,j)$. \textbf{(d)-(e)} Confocal images at different times for systems at 250$\mu$M and 500$\mu$M, respectively. Note that the ones at 500$\mu$M display growing red patches that are themselves filled with green droplets even after several hours. The two sample mixtures were prepared from the same pure batches of DNS-A and of DNS-B, following the protocol in Supplementary Section \ref{SM-sec:expInterpenetration}. \textbf{(f)} Temporal evolution of the intensity of the yellow signal. Each data point represents the mean $\pm$ SD of the yellow intensity across the image (calculated at a fixed time-point). The inset shows the mean $\pm$ SD of yellow intensity for each concentration averaged over all the displayed time points. We also show representative confocal images from experiments in the respective phases, with scale bars of 250 $\mu$m for all images.}
	\label{fig:interpenetration}
\end{figure*}

To summarise, our analysis from simulations suggest that as one increases the concentration of building blocks in limited valence gels, the microstructure displays the formation of longer (and larger) loops but tighter pore sizes. We realised that these two seemingly contrasting behaviours can be reconciled if these networks formed interpenetrating structures. Indeed, interpenetration would reduce the space in between DNA strands, yet require longer loops to accommodate loop threading.
Motivated by this conjecture, we set out to detect the presence of interpenetration in our simulations: we selected all nanostars within different spherical neighbourhoods of our simulations and created adjacency graphs. We discovered that above $\rho^*$ most subsets form two unconnected graphs and show the presence of interpenetrated structures (see Fig.~\ref{fig:interpenetration}(a)). Our evidence strongly suggests that generic limited valence networks form interpenetrating networks, leading us to anticipate the presence of topologically complex motifs. While this picture is similar to the one discovered in the hydrogen bond networks of water~\cite{Sciortino2022} and ice~\cite{Fosado2024ice}, we argue that in DNAns networks these topological motifs may play a role in determining the gel's material properties.

\subsection*{\centering Confocal imaging supports interpenetration in DNAns gels}

To experimentally test interpenetration in DNAns hydrogels, we designed two types of nanostars (DNS-A and DNS-B) that have the same geometry and nucleotides composition but enough variations in their core sequences to avoid misfolding and unintended A-B cross-linking~\cite{Saleh2020} (see Fig.~\ref{fig:interpenetration}(b) and Supplementary Table~\ref{SM-table:seqAB}). The overhang sequences employed for the two designs are self-complementary but distinct for the A and B nanostars, so that only A-A or B-B hybridisation is enthalpically favoured and with equal self-binding strength (see Supplementary Section~\ref{SM-sec:expInterpenetration}). To identify each design, we introduced modified nucleotides bound to fluorophores -- FAMK for DNS-A and Cy3 for DNS-B -- in the middle of one of the dsDNA arms of the nanostars.

Upon mixing equimolar amounts of DNS-A and DNS-B, each beyond their own gel binodal, microemulsions are formed: small A-rich droplets form in a large B-rich droplet, and vice-versa. We multiply the normalised green $I_g$ and red $I_r$ signal intensities in each pixel to obtain a two-dimensional map of ``yellow'' pixels with intensity, $I_y(i,j) = I_r(i,j) \times I_g(i,j)$ (Fig.~\ref{fig:interpenetration}(c)). A large value of $I_y$ indicates the presence of both DNS-A and DNS-B, which we identify as an interpenetrated region.

We tracked the temporal evolution of confocal images from experiments at two different nanostar concentrations. At $C=250$$\mu$M, within the phase-separation region, samples are observed to completely demix over time (Fig.~\ref{fig:interpenetration}(d,f)). Beyond the phase-separation region (at $C=500$$\mu$M), we observed that mixed ``yellow'' regions are stable over hours and even weeks (Fig.~\ref{fig:interpenetration}(e,f) and Supplementary Fig.~\ref{SM-fig:Matrix}). To quantify the extent of interpenetration as a function of DNAns concentration, we computed the average intensity of $I_y(i,j)$ over time using more than 20 representative images taken at long times, to obtain $\langle I_{y} \rangle$. In the phase separation region, we observed fully demixed phases and $\langle I_{y} \rangle \simeq 0$ (Fig.~\ref{fig:interpenetration}(f)). Using the same microscope settings, gel samples show a significantly larger value, on average $\langle I_{y} \rangle \simeq 0.4$. These observations suggest the presence of partially interpenetrated structures in DNAns hydrogels, in line with indirect evidence from dynamic light scattering obtained in a different DNAns hydrogel design~\cite{Sciortino2022Interp}. 

We note that while yellow regions (demarcating high local density of both DNS-A and DNS-B) confirm the interpenetration between the two networks, darker regions (with an unbalanced density of DNS-A and DNS-B) may still form interpenetrated structures made by mainly one type of DNAns. Therefore, we have verified (see Supplementary Fig.~\ref{SM-fig:SERvsTPR}) that the intensity of the individual signals, either red or green, is lower (higher) in the balanced (unbalanced) regions. We have also confirmed through MD simulations that a binary mixture of DNAns yield the same interpenetrated structures and scaling of elasticity than the monodisperse system (see Supplementary Fig.~\ref{SM-fig:mixedAB}).

\begin{figure*}[t!]
	\centering
	\includegraphics[width=0.85\textwidth]{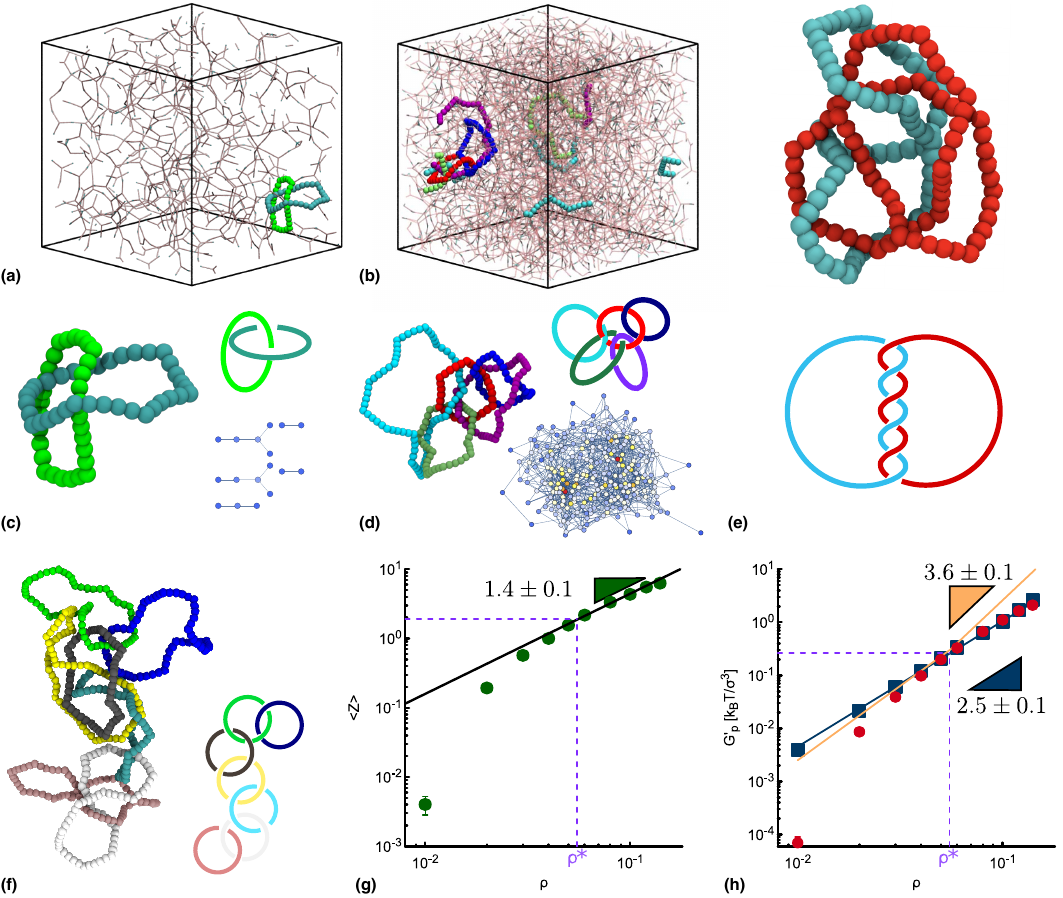}
	\caption{\textbf{Topological Elasticity.} \textbf{(a)-(b)} Representative snapshots from simulations of networks at $\rho=0.02$ (with $N_{minloop}=47$ and $\mathcal{L}=12$) and $\rho=0.1$ (with $N_{minloop}=256$ and $\mathcal{L}=1159$), respectively. \textbf{(c)-(d)} Example of linked rings found in the snapshots from (a)-(b). Network diagrams represent minimum loops as nodes and linkages between loops as edges (unlinked minimum loops not shown). \textbf{(e-f)} At high concentration ($\rho=0.1$), more complex topologies are observed: \textbf{(e)} minimum loops with linking number 3 and \textbf{(f)} polycatenanes. \textbf{(g)} Average number of links per minimum loop, i.e., linking valence, as a function of concentration. Error bars are smaller than symbols. \textbf{(h)} Scaling of elasticity with concentration comparing Green-Kubo simulations (dark blue, with $G^{\prime}\sim\rho^{2.5}$ in the whole range of concentrations, see also Fig.~\ref{fig:experiments}(i)) and predictions from (i) mesh size (yellow line) and (ii) linking number (red  circles). The last two quantities have been multiplied by 0.03 and 0.001, respectively, to help the comparison of the scaling. In (g) and (h) values are expressed as mean $\pm$ SEM, obtained from at least 25 independent configurations per volume fraction.}
	\label{fig:NetworkTopology}
\end{figure*}

\subsection*{\centering Characterizing topology motifs in interpenetrated networks}

Having found strong evidence for the emergence of interpenetrated structures in both simulations and experiments, we decided to characterise the ensuing topological motifs. Typical configurations of the networks formed at two volume fractions, $\rho=0.02$ and $\rho=0.1$, are shown in Figs.~\ref{fig:NetworkTopology}(a) and (b), where it is clear that as the concentration of DNAns increases, topological links between the minimum loops appear to be more abundant in the network.

We then computed the Gauss linking number between all possible pairs of minimum loops in the network as:
\begin{equation}
\text{Lk}(\gamma_{i},\gamma^\prime_{j})=\oint_{\gamma_{i}} \oint_{\gamma^\prime_{j}} ds \, ds^\prime \dfrac{(\bm{t}(s) \times \bm{t}(s^\prime)) \cdot (\bm{r}(s) - \bm{r}(s^\prime))}{|\bm{r}(s) - \bm{r}(s^\prime)|^3},
\label{eq:lk}
\end{equation}
where $\gamma_{i}$ and $\gamma^\prime_{j}$ are loops formed by fragments of DNAns, $\bm{t}(s)$ is the tangent at position $s$ along the loop and $\bm{r}(s)$ the 3D coordinate of the segment at position $s$ along the loop. Strikingly, we discovered that the fraction of minimum loops that are linked (at least once) grows from $0$ at $\rho=0.01$ to $95\%$ at the overlap concentration $\rho^{*}=0.056$. Additionally, at large concentrations, we observed the emergence of complex topologies, such as multiply-linked loops (Fig.~\ref{fig:NetworkTopology}(e)) and polycatenanes (Fig.~\ref{fig:NetworkTopology}(f)). Fascinatingly, these interlinked loops themselves form a percolating network (Fig.~\ref{fig:NetworkTopology}(d), inset), thus establishing a network-within-network structure in DNAns gels. 

Since DNA nanostars interact through physical (reversible) bonds, the minimum loops are not permanent and we observed that the system's topology changes in time. In our simulations, the networks reach a steady state in which the total number of minimum loops and total linking ($\mathcal{L}=\sum_{i>j}^{N_{minloops}} \lvert \text{Lk}(i,j) \rvert$) fluctuate around constant values (see Supplementary Figs.~\ref{SM-fig:minloop}(e) and \ref{SM-fig:minloopLk}(c)). The average linking number per (minimum) loop, i.e., the average linking valence $\langle \text{Z} \rangle = \mathcal{L}/N_{minloop}$, as a function of the concentration of DNAns, is shown in Fig.~\ref{fig:NetworkTopology}(g). Interestingly, the overlap concentration $\rho^{*}$ marks the point at which the system undergoes a change in its topological state: at $\rho<\rho^{*}$ the linking between loops is negligible ($\langle \text{Z} \rangle < 1$), while for $\rho \geq \rho^{*}$ all loops are on average linked at least once ($\langle \text{Z} \rangle > 1$). We also note that the average linking valence grows as $\langle \text{Z} \rangle \sim \rho^{1.4}$, which is consistent with a simple geometrical consideration on the scaling of the number of overlapping minimum loops, expected to follow $\langle Z \rangle \sim N_{minloop} R^3_{g,minloop}/V \sim \rho \langle l_{min}\rangle^3 \sim \rho^{1.3}$ (where we used $\langle l_{min} \rangle \sim \rho^{0.1}$, Fig.~\ref{fig:NetworkGeometry}(f)).
 
This topological transition in the linking of (minimum) loops at $\rho^*$ explains the counter intuitive results seen above: as the concentration of DNAns increases, not only more loops appear in the network but these loops become larger, concomitantly with the emergence of more tightly packed, interpenetrated structures, which in turn favour the formation of links and increase the topological complexity of the network.

We argue that at timescales shorter than the typical unbinding time of individual DNAns (in our DNAns design $\tau_u \simeq 1$ s at 25$^\circ$C), perturbations to the network will effectively apply stress to these topological interlocked features as if the network topology was frozen. Therefore, the topology of interlinked minimum loops would determine the elastic response of the gel at frequencies $\omega > \tau_u^{-1}$. Given that our elastic behaviour was measured around $\omega \simeq 100$ Hz $\simeq 10$ $\tau_u^{-1}$ (Fig.~\ref{fig:experiments}), we expect the scaling of $G^\prime_p$ to be determined by linking of the minimum loops.

\subsection*{\centering Topological elasticity}

The elasticity of polymer networks can be understood as the combined effect of the density of crosslinks (related to the network connectivity) and that of entanglements~\cite{Rubenstein2022,Everarers2008}. At high crosslinking density, crosslinks dominate over entanglements and  classic theories, such as affine and phantom network models, neglect excluded volume interactions between polymers and the presence of entanglements~\cite{Hubert1947PhanNet}. 
Indeed, for building blocks with limited valence $f$, phantom network theory predicts $G^{\prime}_{p}\sim \rho (f-2)/f$.
Simulations of disordered phantom networks showed that this simple relation fails to predict the elasticity of the system which is instead better described by an even weaker scaling $G^{\prime}_{p} \sim \rho^{1/3}$~\cite{Sorichetti2021}. Thus, current models fail to predict the non-linear scaling of the elastic plateau with concentration observed in our simulations and experiments $G^{\prime}_{p} \sim \rho^{2.5}$ (Fig.~\ref{fig:experiments}(e,i)). Large scaling exponents for the elastic behaviour at large frequencies were also found in previous works on similar complex fluids made of limited valence building blocks~\cite{ConradPNAS2019, Sorichetti2023}. A comprehensive and predictive theory for the behaviour of limited valence reversible gels, should account for the topological features we uncovered and reported in the previous sections.
Specifically, we note that for $\rho \geq \rho^{*}$, if the elasticity at high frequency was determined by the mesh size, we would expect it to phenomenologically scale as (see Supplementary Section~\ref{SM-sec:overlapConc})
\begin{equation}
G^{\prime}_{p} \sim \dfrac{k_B T N_{minloop}}{\xi^3} \sim \rho^{3.6} \, ,
\end{equation}
where we made use of the fits $N_{minloop} \sim \rho$ and $\xi \sim \rho^{-0.84}$ at large $\rho$ (see Fig.~\ref{fig:NetworkGeometry}). This scaling is shown in Fig.~\ref{fig:NetworkTopology}(h) to significantly deviate from direct Green-Kubo (GK) measurements at large $\rho$ (Fig.~\ref{fig:NetworkTopology}(h), blue squares).

Instead, we find that the scaling of the GK-measured elasticity is better explained by the following simple argument: assuming that linking between minimum loops becomes the dominant entanglement mechanism on timescales shorter than the typical melting time of the sticky ends, the high frequency elastic plateau should be proportional to (i) the density of elastically active polymers $\rho_e$ (the density of minimum loops) and (ii) the number of entanglements per polymer $Z_e$ (the average linking valence). We thus expect 
\begin{equation}
G^{\prime}_{p} \sim \rho_{e} \text{Z}_e \sim  N_{minloop} \langle \text{Z} \rangle \sim \mathcal{L} \sim \rho^{2.4} \, ,
\end{equation}
where we made use of $N_{minloop} \sim \rho$ and $\langle \text{Z} \rangle \sim \rho^{1.4}$ (see Fig.~\ref{fig:NetworkTopology}(g)). This scaling is in very good agreement with both experiments and simulations (Fig.~\ref{fig:NetworkTopology}(h)). The small discrepancy in the predicted exponent (2.4) and the one measured through GK calculations (2.5) is within uncertainty. We note that this discrepancy could also be attributed to the fact that our topological analysis neglects links between second-order loops (larger than the minimum ones), and loop threading.

\subsection*{\centering Conclusions} 

In this paper, we have tackled a long-standing question: the connection between a material's microscopic topology and its macroscopic material properties. We focused on soft viscoelastic gels made of limited valence building blocks with transient, non-covalent crosslinks and realised them in the lab using tri-armed DNA nanostars.

The central conclusion drawn from our findings is that at timescales shorter than DNAns sticky-end unbinding, the elastic response of the gel is dominated by linked loops, akin to Olympic networks~\cite{degennes_scaling,Krajina2018} like kinetoplast DNA~\cite{Michieletto2014kdna,Klotz2020,He2022}. In fact, the linking of minimum loops grows as $\mathcal{L} \sim N_{minloops} \langle Z \rangle \sim \rho^{2.4}$, matching the scaling of the elastic plateau (Fig.~\ref{fig:NetworkTopology}). The simple linear relationship found ($G^\prime_{p} \sim \mathcal{L}$) is very powerful as it connects the rheology of the network with its topology (in the formal mathematical sense of the term, not referring to generic entanglements).

Based on our findings, we argue that if we were to design DNAns with longer arms, or to use binary mixtures of DNAns with different arms' length, we would obtain stronger networks, despite the effective reduction of cross-links per unit volume. Additionally, by designing DNAns with longer sticky ends (and hence longer unbinding times) we would realise semi-irreversible interactions, and complex fluids with an elastic behaviour at lower frequencies. In fact, through DNA ligation we could render DNAns-DNAns bonds irreversible. Whether the ensuing low frequency elastic behaviour could be described by the linkages between minimum loops in the network is to be determined, as the gelation kinetics may also contribute to affect the scaling.

We expect that our results should hold for other network-forming structures made by limited valence building blocks, as they favour the formation of complex porous networks. We expect that the ``topological elasticity'' mechanism uncovered in this paper will be relevant to understand the physical properties of generic networks covering different time and length scales and made by, for instance, patchy colloids~\cite{Sciortino2017,Sorichetti2023}, polymers~\cite{Zhong2016}, molecular liquids such as water and ice~\cite{Sciortino2022,Fosado2024ice}, DNA~\cite{SciortinoPNAS2013,EiserPNAS2018,ConradPNAS2019}, polymer rings~\cite{OConnor2020,Michieletto2016pnas,Krajina2018,Rauscher2020}, and potentially other complex fluids such as polymer gels and liquid crystals with entangled defects~\cite{Dudukovic2014, Wood2011}. Indeed, topology is by its nature universally found across physical and biological systems. 
Overall, our findings contribute to a better fundamental understanding of how formal topological motifs within a network affect the material properties of the bulk, and will also guide the design of topologically-controlled structural and mechanical properties of generic network-forming liquids and materials.

\section*{\centering Acknowledgements}
DM acknowledges the Royal Society and the European Research Council (grant agreement No 947918, TAP) for funding. The authors also acknowledge the contribution of the COST Action Eutopia, CA17139. The authors would like to thank the referees Prof Omar Saleh and Prof Emanuela Del Gado for insightful comments on our work and a scientifically rewarding review process. For the purpose of open access, the authors have applied a Creative Commons Attribution (CCBY) licence to any Author Accepted Manuscript version arising from this submission.

\section*{\centering Author contributions}
YF, DM and GP designed the research. YF conducted simulations, developed analysis codes, and performed analysis from simulations. SW performed BR experiments. GP performed MR and confocal imaging. GP and SW analyzed data from experiments. DM and YF supervised research and wrote the original draft. All authors reviewed the manuscript. DM acquired funding.

\section*{\centering Competing interests}
The authors declare no competing interests.

\bibliographystyle{naturemag}
\bibliography{TNENLV}

\section*{\centering Methods}
\subsection*{\centering DNA nanostar design}
Nanostar motifs are assembled from three single-stranded oligonucleotides, 49 nucleotides long each and with sequences reported in Supplementary Table~\ref{SM-table:seq}. These motifs were designed using NUPACK~\citemethods{NUPACK}, with minor modifications from those originally proposed in Ref.~\cite{BiffiSM2015}. Each double-stranded DNA arm is 20 base-pairs (bp) long, and terminates in a self-complementary 6-nucleotide fragment with sequence 5'-CGATCG-3'. This sticky end is equal for all three arms, allowing the non-specific hybridization of two nanostars (ns): any of the three arms of one ns can hybridize with any (but only one) of the arms of another ns. Unpaired adenines (A) are introduced at the core of the Y-shaped structure and before the sticky end to enhance the internal flexibility of the nanostar and the nanostar–nanostar bond. The nucleotide sequences for the three-armed nanostar were acquired from Integrated DNA Technologies (IDT, https://www.idtdna.com/pages). Oligonucleotide tubes were then dissolved in UltraPure DNase/RNase-Free Distilled Water (Invitrogen, 11538646) and used to form nanostar samples as described in Supplementary Section \ref{SM-sec:hydrogelprep}. Nanostar samples were prepared at the desired concentration in Nanostar Buffer, containing the following reagents: Tris (Invitrogen, 15504020), sodium acetate (Sigma-Aldrich, 33209-M), EDTA (Sigma-Aldrich, ED2P), and NaCl (Thermo Scientific Chemicals, 10092740).

\subsection*{\centering Microrheology}
We used particle tracking microrheology to assess the mechanical properties of samples of nanostras at different concentrations. Polystyrene beads of 200nm size (Sigma-Aldrich, 69057) are spiked in the solution and tracked by using trackpy (\url{github. com/soft-matter/trackpy}). The viscoelastic properties of the sample are then obtained by applying the generalized Stokes-Einstein relation~\citemethods{Mason1995} to the mean squared displacement of the beads (see Supplementary Section~\ref{SM-sec:MR} and Supplementary Fig.~\ref{SM-fig:MR} for details).

\subsection*{\centering Bulk rheology}
Frequency sweeps were performed at a shear strain of $\gamma= 0.5\%$ within the linear viscoelastic region. DNAns solutions behave as near-Maxwellian viscoelastic fluids, with low-frequency liquid behaviour ($G^{\prime\prime} > G^{\prime}$, with $G^{\prime} \sim \omega^{2}$ and $G^{\prime\prime}\sim \omega$) separated by a crossover frequency ($\omega_{0}$), from high-frequency solid-like behaviour ($G^{\prime} > G^{\prime\prime}$) and with a plateau modulus $G^{\prime}_{p}$ (see Supplementary Section~\ref{SM-sec:BR} and Supplementary Fig.~\ref{SM-fig:BR} for details).

\subsection*{\centering Confocal imaging}
DNS-A and DNS-B solutions are prepared in two separate test tubes to prevent the formation of unintended secondary structures during the annealing step. In both cases, the fluorescently tagged and untagged oligomers (purchased from IDT, with sequences reported in Supplementary Table~\ref{SM-table:seqAB}) are mixed at a molar ratio of 1:10 (or 1:20). Samples are visualised with a Zeiss LSM700 confocal microscope with a scan time of 15.49s and a laser intensity equal to 2.4\% (no bleaching effects were detected at this intensity value). During the image acquisition, the 488nm and the 555nm lasers scan the sample sequentially to excite FAMK(DNS-A, green) and Cy3(DNS-B, red). To visualise the final image, we assemble the two channels into one via Fiji (see Supplementary Section~\ref{SM-sec:expInterpenetration} and Supplementary Fig.~\ref{SM-fig:Matrix} for details).

\subsection*{\centering Coarse-grained model and simulations}
In simulations, seven beads are assembled into a Y-shaped rigid structure that resembles a nanostar. Small patches are placed at the end of each arm to represent sticky sites. The core particles are purely repulsive, with an excluded volume of $\sigma = 2.5$ nm $\simeq 8$ base-pairs. The patches interact via a short-range attractive Morse potential (see Eq.~\ref{SM-eq:morse} and Supplementary Section~\ref{SM-sec:simDetails}), and the geometry of the bead-patch and parameters of the Morse potential are set to ensure one-to-one binding of the simulated nanostars, thereby mimicking sticky ends hybridisation. Molecular dynamics simulations were performed in the NVT ensemble using LAMMPS, with implicit solvent (Langevin dynamics). The integration timestep was set to $dt = 0.01\tau_{Br}$ ($\tau_{Br}=k_{B}T/\gamma$ is the Brownian time and $\gamma$ is the friction, set to 1 in LJ units) and the temperature to $T=1\epsilon/k_{B}$. Initial configurations are prepared by performing first an equilibration run for $5\times10^{5} \tau_{Br}$ while attraction between DNAns is not allowed. Then, we turned-on the Morse attraction and perform a production run for $10^{6}\tau_{Br}$. We identified hybridization between adjacent nanostars when their patches are located at a distance $r\leq 0.2\sigma$ (the cut-off distance of the attraction between patches).

Elasticity was measured in simulations first by mimicking MR experiments in which we embedded a large bead in the sample and computed the $G^{\prime}_{p}$ from the mean squared displacement of the bead (see Supplementary Fig.~\ref{SM-fig:MR}). We also performed Green-Kubo simulations and find the elastic plateau from the autocorrelation of the stress tensor (see Supplementary Section~\ref{SM-sec:elastSim} and Supplementary Fig. \ref{SM-fig:GK}).

\section*{\centering Data availability}
Datasets created during the current study and scripts to generate the figures in this article are openly available in~\citemethods{repoFosado2024}. Data is also deposited at:\\
\href{https://git.ecdf.ed.ac.uk/ygutier2/data-j1Topo}{https://git.ecdf.ed.ac.uk/ygutier2/data-j1Topo}

\section*{\centering Code availability}
Codes to perform simulations and topology analysis are openly available in~\citemethods{repoFosado2024} and at: \href{https://git.ecdf.ed.ac.uk/ygutier2/j1Topo}{https://git.ecdf.ed.ac.uk/ygutier2/j1Topo}  

\bibliographystylemethods{naturemag}
\bibliographymethods{methods}

\end{document}